\documentclass[sigplan,10pt]{acmart}
\usepackage{booktabs}   
\usepackage{subcaption} 

\usepackage{xcolor}     
\definecolor{dkgreen}{rgb}{0,0.5,0}
\definecolor{green}{rgb}{0,0.5,0}
\definecolor{gray}{rgb}{0.5,0.5,0.5}
\definecolor{mauve}{rgb}{0.58,0,0.52}

\usepackage{algorithm}
\usepackage[noend]{algpseudocode}
\usepackage{listings}
\usepackage{mathtools}
\usepackage{enumitem}
\lstdefinestyle{code}{
  language=Java,
  showstringspaces=false,
  columns=flexible,
  basicstyle={\ttfamily\footnotesize},
  numbers=left,
  numberstyle=\scriptsize\color{black},
  keywordstyle=\ttfamily\color{mauve},
  numbersep=5pt,
  commentstyle=\color{dkgreen},
  stringstyle=\color{mauve},
  breaklines=true,
  breakatwhitespace=false,
  tabsize=2,
  xleftmargin=0.4cm,
  framexleftmargin=1.5em,
  framesep=2pt,
  escapechar=|,
}
\def\tool{\textsc{AutoTornado}}
\def\parcons{\texttt{@Parallel}}

\acmConference[arXiv]{Preprint submitted to arXiv}{May}{2022}
\acmYear{2022}
\acmISBN{} 
\acmDOI{} 
\startPage{1}

\setcopyright{none}

\begin{document}

\title[\tool{}]{Can We Run in Parallel?}         
\subtitle{Automating Loop Parallelization for TornadoVM}                     


\author{Rishi Sharma}
\authornote{Work performed during the author's stay at IIT Mandi, India.}          
\affiliation{
  \institution{IIT Mandi}            
  \country{India}                    
}
\email{rishi-sharma@outlook.com}          

\author{Shreyansh Kulshreshtha}
\authornote{Work performed during the author's stay at IIT Mandi, India.}          
\affiliation{
  \institution{IIT Mandi}           
  \country{India}                   
}
\email{shreyanshkuls@outlook.com}         

\author{Manas Thakur}
\affiliation{
  \institution{IIT Mandi}           
  \country{India}                    
}
\email{manas@iitmandi.ac.in}          

\begin{abstract}
    With the advent of multi-core systems, GPUs and FPGAs, loop parallelization
    has become a promising way to speed-up program execution.
    In order to stay up with time, various performance-oriented programming
    languages provide a multitude of constructs to allow programmers to write
    parallelizable loops.
    Correspondingly, researchers have developed techniques to automatically
    parallelize loops that do not carry dependences across iterations, and/or
    call pure functions.
    However, in managed languages with platform-independent runtimes such as
    Java, it is practically infeasible to perform complex dependence analysis
    during JIT compilation.
    In this paper, we propose \tool{}, a first of its kind static+JIT loop
    parallelizer for Java programs that parallelizes loops for heterogeneous
    architectures using TornadoVM (a Graal-based VM that supports insertion of
    \parcons{} constructs for loop parallelization).

    \tool{} performs sophisticated dependence and purity analysis of Java
    programs statically, in the Soot framework, to generate constraints
    encoding conditions under which a given loop can be parallelized.
    The generated constraints are then fed to the Z3 theorem prover (which we
    have integrated with Soot) to annotate canonical {\tt for} loops that can
    be parallelized using the \parcons{} construct.
    We have also added runtime support in TornadoVM to use static analysis
    results for loop parallelization.
    Our evaluation over several standard parallelization kernels shows that
    \tool{} correctly parallelizes 61.3\% of manually parallelizable loops,
    with an efficient static analysis and a near-zero runtime overhead.
    To the best of our knowledge, \tool{} is not only the first tool that
    performs program-analysis based parallelization for a real-world JVM, but
    also the first to integrate Z3 with Soot for loop parallelization.
\end{abstract}



\maketitle

\section{Introduction}
\label{s:intro}

With the onset of multicore and heterogeneous systems over the last two decades, several programming languages were enriched with various ways to write concurrent programs that could reap out the benefits of the available hardware.
Few languages provide inbuilt facilities to fork and launch multiple threads, whereas others support writing concurrent programs with extensions or libraries.
At program level, writing complex computations invariably involves iterating over large data sets in loops.
However, languages such as Java, though provide an in-built ability to write multithreaded programs, do not allow the programmer to directly mark loops for parallelization.
One of the useful developments in this space has been the design of TornadoVM~\cite{tornadovm}.

TornadoVM is a Java virtual machine (JVM) that extends OpenJDK and GraalVM~\cite{graal} with a facility to parallelize {\tt for} loops across heterogeneous architectures.
TornadoVM allows programmers to annotate {\tt for} loops in Java source code with an {\tt @Parallel} construct, preserves the annotations in Java Bytecode, and then parallelizes the marked loops on the available hardware (CPUs, GPUs as well as FPGAs), during execution in the virtual machine.
This not only alleviates the need to write different parallelization back-ends for different architectural components, but also enriches Java with a facility to express parallelization opportunities at loop-level.
However, identifying which loops can be parallelized is a non-trivial problem and involves sophisticated program analyses for even trivial programs (those involving accesses to arrays with indices being affine functions of loop variables).

\begin{figure}
\begin{lstlisting}
int elem_sq(int x) { return x*x; }
int array_sq(int[] x) {
    for (int i = 0; i < x.length; i++) {
        int t = elem_sq(x[i]);
        x[i] = t; }
    return x; }
\end{lstlisting}
\caption{A Java code snippet to demonstrate the analyses required for loop parallelization.}
\label{fig:intro}
\end{figure}

As an example, consider the Java code snippet shown in Figure~\ref{fig:intro}.
In order to determine if the {\tt for} loop in function {\tt array\_sq} can be parallelized without changing the semantics of the program, we need to (i) extract the array indices at statements 4 and 5; (ii) find out if the indices may access the same location in a conflicting manner across iterations; and (iii) check if the call to the function {\tt elem\_sq} may cause any side-effect.
This translates to performing dependence analysis to find constraints under which a loop can be parallelized, solving the identified constraints (possibly using a constraint solver), pointer analysis to identify aliases and to resolve method calls, and an interprocedural purity analysis in the VM.
However, performing multiple such precise interprocedural analyses in a JVM, where the program-analysis time directly affects the execution time of a program, is prohibitively expensive.
In this paper, we present a tool named \tool{} that addresses all the challenges listed above: it performs precise analysis of Java bytecode and marks loops for parallelization in TornadoVM, with negligible overhead during program execution.

\tool{} first performs dependence analysis over Java programs, in the Soot framework~\cite{soot}.
The dependence analysis generates constraints (over iteration variables) that encode conditions under which a given candidate loop can be parallelized.
\tool{} then feeds these constraints to the Z3~\cite{z3} theorem prover (which we have integrated with Soot), to determine if the constraints are satisfiable.
In case the loop additionally makes some function calls, \tool{} also analyzes the called function(s) for purity.
Finally, if there are no dependences found by Z3, and if the loop does not call any impure functions, then the given loop can be marked for parallelization.
Inspired by the PYE framework~\cite{pye} (which proposes ways to interface analyses across static and JIT compilers), \tool{} stores and conveys information about such parallelizable loops to our modified TornadoVM, which simply uses \tool{}-results to parallelize the identified loops during program execution.

The key contribution of our approach is to enable complex dependence- and purity-analysis based automatic loop parallelization for a VM, without spending much time in the VM.
To evaluate the efficacy of this approach, we compared the precision of \tool{}-identified parallelization opportunities with manually marked loops, for 14 benchmarks from a Java version of the PolyBench suite~\cite{polybench} (part of the TornadoVM repository).
We found that \tool{} successfully marks 61.3\% of loops for parallelization, and even identifies few loops that were not identified manually.
We also measured the speed-ups achieved due to \tool{}-identified parallel loops, and found that it is significant (3.99x, on average, across all the benchmarks under consideration).
In order to evaluate the trade-off between storing additional static-analysis results and performing expensive analyses in the VM, we computed the space overhead of our result files and the analysis time spent in Soot+Z3.
The results show that \tool{} enables loop parallelization in TornadoVM with a small storage overhead (10.2\% of class files, which can be further reduced by adding code annotations in class files themselves), negligible run-time overhead (2.85\% of the execution-time of benchmarks), and that performing those analyses in the VM would have been prohibitively expensive (57.58\% of the total execution-time itself).

To the best of our knowledge, \tool{} is not only the first tool that performs program-analysis based parallelization for a real-world JVM, but is also the first to integrate Soot with Z3 for loop parallelization.
The purity analysis on its own is a Soot enhancement for recent Java versions and is usable as an add-on by other analyses and frameworks.
Additionally, the idea of carrying static-analysis results to a Java runtime in itself is novel enough and presents interesting engineering challenges; this manuscript traverses our exploration of the design space and describes our solutions, throughout the presentation.
We thus believe that in the vast research span that focuses on parallelization, our work is a significant step in solving the associated challenges for languages with managed runtimes such as Java.

The rest of the paper is organized as follows.
Section~\ref{s:bg} introduces the tools and analyses that we use in our work.
Section~\ref{s:analysis} gives an overview of the architecture of our proposed approach (\tool{}) and elaborates on each of the analysis components.
Section~\ref{s:runtime} describes the changes that we do in TornadoVM to support reading static-analysis results, and Section~\ref{s:discsn} presents few design choices that we made.
Section~\ref{s:eval} evaluates our tool with respect to the precision of \tool{}-marked annotations, the overheads in terms of storage and runtime, and the speed-ups achieved with \tool{}-parallelized loops.
Finally, Section~\ref{s:related} discusses few related works and Section~\ref{s:concl} concludes the paper.

\section{Background}
\label{s:bg}

This section introduces few preliminary concepts, tools, and analyses that are used throughout the paper.

\begin{figure}[ht!] 
    \begin{center}
        \begin{tabular}{c}
\begin{lstlisting}
public void swap(int[] ar) {
    int n = ar.length;
    for (int i = 0; i < n; i++) {
        int temp = ar[i];
        ar[i] = ar[i-1];
        ar[i-1] = temp; } }
\end{lstlisting}\hspace*{\fill}
        \end{tabular}
    \end{center}
    \caption{Code to swap adjacent array elements.}
    \label{fig:depCode0}
\end{figure}

{\bf 1. Loop-level dependencies and parallelization.}
Figure~\ref{fig:depCode0} shows a code where each iteration swaps adjacent array elements by using a local temporary variable \texttt{temp}. \textit{Read-after-write} or \textit{write-after-read} dependence occurs when an iteration of the loop reads from a memory location and another iteration tries to write to the same memory location. Write-after-write dependence occurs when two separate loop iterations write to the same memory location. The write to \texttt{ar[i]} from iteration \texttt{i} and the read of \texttt{ar[i-1]} from iteration \texttt{i+1} on line $5$ imply a \textit{read-after-write} dependence. Similarly, the writes to \texttt{ar[i]} on line $5$ from iteration \texttt{i} and \texttt{ar[i-1]} on line $6$ from iteration \texttt{i+1} imply a \textit{write-after-write} dependence. As the memory location is mutated in both of the cases, these dependences can lead to incorrect results if the iterations are run in parallel.
There is another benign dependence that we can ignore: \textit{read-after-read} dependence, where two different iterations read from the same memory location. As no memory location is mutated, parallelization poses no threat.
Apart from not having non-benign dependencies, in order to be parallelizable, a loop must not perform any operation (either directly or through a function call) that may lead to side-effects.


{\bf 2. Soot}~\cite{soot} is a popular framework for analyzing Java bytecode. The framework is written in Java and provides different intermediate representations (IRs) for representing the Java code. We use the Jimple IR for analysis, which is a typed three-address representation. We also use Soot's points-to analysis Spark~\cite{spark} for constructing call-graphs and for deriving alias relationships among references.

{\bf 3. TornadoVM}~\cite{tornadovm} is a plug-in to OpenJDK and GraalVM (standard Java runtime environments) that allows programmers to automatically run Java programs on heterogeneous hardware. It works by reconfiguring applications, at runtime, for hardware acceleration based on the currently available hardware resources. TornadoVM currently targets OpenCL-compatible devices and executes code on multi-core CPUs, dedicated and integrated GPUs, and FPGAs.
TornadoVM can parallelize canonical \texttt{for} loops whose index variable and increment operation follow certain conditions; the loops to be parallelized in turn need to be annotated as \texttt{@Parallel}.

%

{\bf 4. Z3}~\cite{z3} is an SMT solver and theorem prover developed by Microsoft Research. SMT or Satisfiability Modulo Theories problems are decision problems for logical formulae with respect to combinations of theories such as arithmetic, bit-vectors, arrays, and uninterpreted functions. Z3, being a very efficient SMT solver, is widely used for solving problems that arise in software verification and analysis.

\section{\tool{}}
\label{s:analysis}

\begin{figure*}
    \includegraphics[scale=0.67]{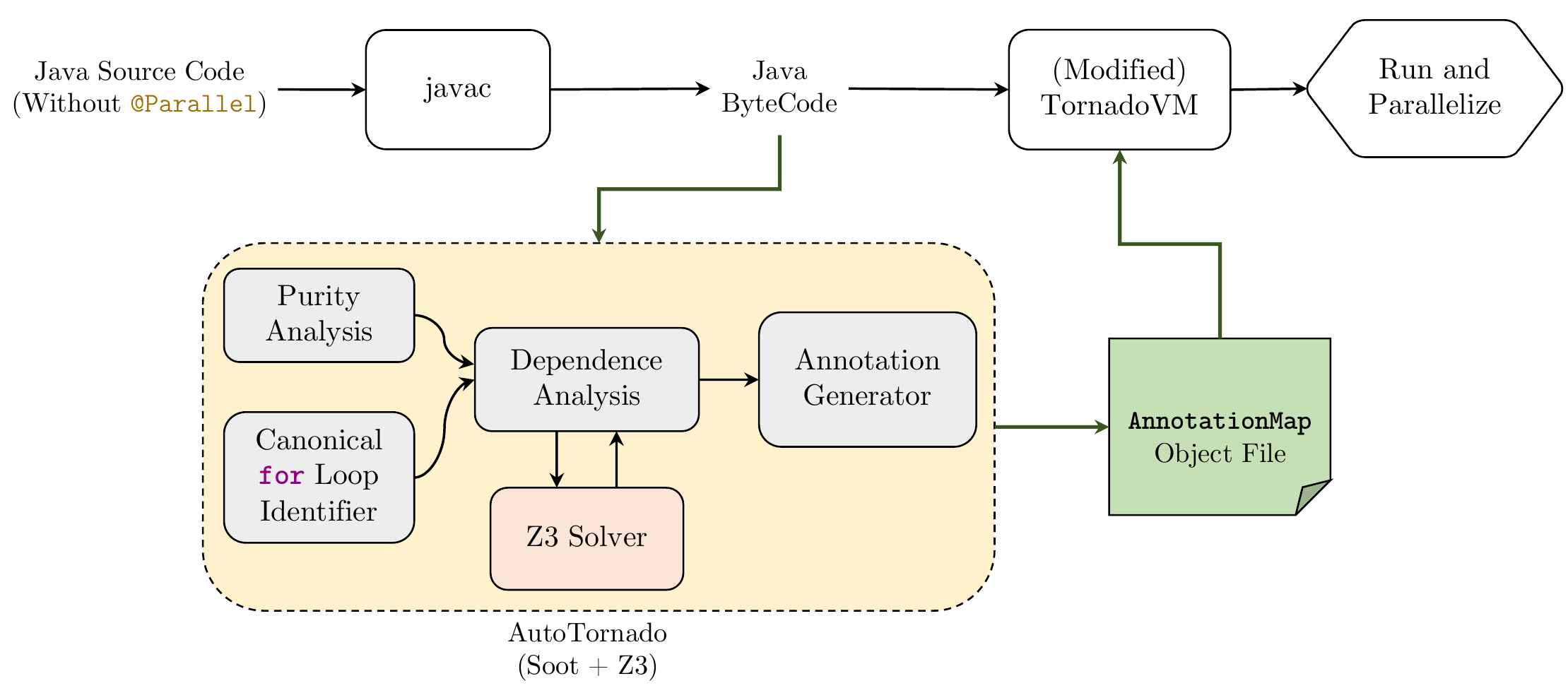}
    \caption{Architecture of \tool{}.}
\label{fig:block-diagram}
\end{figure*}

In this section, we first  illustrate the architecture of \tool{}, and then describe each of its modules.
While describing each module, we first motivate the need for the module, followed by our proposed solution.

Figure~\ref{fig:block-diagram} shows how \tool{} identifies parallelizable loops and conveys this information to {\tt TornadoVM}.
Given Java bytecode, \tool{} first identifies canonical {\tt for} loops (candidates for parallelization) and then performs dependence analysis over the loop bodies, along with purity analysis over called functions.
The dependence constraints are fed to the Z3 theorem prover, which helps in classifying which loops can be parallelized without changing the underlying semantics.
Finally, \tool{} generates annotations that can be supplied to the {\tt TornadoVM} runtime and adds support to the VM (see Section~\ref{s:runtime}) to parallelize loops identified by our static analysis during execution.

\subsection{Identifying Canonical Loops}
\label{ss:canonical}

Java code is first compiled to Java bytecode which then runs on the JVM. Soot accepts this bytecode as input and converts it into Jimple. However, Java bytecode and Jimple do not have syntactic constructs like \texttt{for} loops. Figure \ref{f:jimple} shows a simple {\tt for} loop that doubles array elements in Jimple representation. As can be observed, identifying the loop, iteration variable, lower-bound, upper-bound, and increment poses a challenge in the Jimple representation.

\begin{figure}[htb!]
    \Description[For loop in Jimple]{A preview of a for loop in jimple comprising of labels and goto statement.}
    \begin{center}
    \begin{tabular}{c}
    {
\begin{lstlisting}
label1:
        if l2 >= 5 goto label2;
        $stack7 = $stack3[l2];
        $stack5 = $stack3[l2];
        $stack6 = 2.0F * $stack5;
        $stack8 = $stack7 + $stack6;
        $stack3[l2] = $stack8;
        l2 = l2 + 1;
        goto label1;
label2:
        return;
\end{lstlisting}}
    \end{tabular}
    {\renewcommand{\baselinestretch}{0.77} \caption{\label{f:jimple}Jimple representation of a simple {\tt for} loop.}}
    \end{center}
\end{figure}

We standardize the domain of loops that we handle to loops identified as {\em canonical} according to the Algorithm \ref{algo:loop}. The loops identified have the following properties: (a) constant lower bound; (b) private iteration variable; (c) linear, positive and constant increment to the iteration variable; (d) condition of the form $i \{<|\le|>|\ge\} u$, (e) as canonical loops, (f) iteration variable only modified in the update statement, and (g) single exit.

\begin{algorithm}
    \footnotesize
    \caption{\label{algo:loop}{Algorithm to identify canonical loops.}}
    \begin{algorithmic}[1]
    \Procedure{isCanonical}{$l$}       
        \State $head \leftarrow l.getHead()$
        \State $cond \leftarrow l.getCondition()$
        \State $back \leftarrow l.backJumpStatement()$
        \State $upd \leftarrow back.predecessor()$
        \State $init \leftarrow head.predecessor()$
        \State $lb \leftarrow init.rhs()$
        \If{$back \neq l.statements.last()$} \Comment{Backjump should be last}
            \State Return $false$
        \ElsIf{$upd.type() \neq Assignment$}
            \State Return $false$
        \EndIf
        \State $iter \leftarrow upd.def()$ \Comment{The variable updated}
        \If{$iter.type() \neq Local$}
            \State Return $false$
        \EndIf
        \If{$init.type() \neq Assignments ~OR ~init.def \neq iter$}
            \State Return $false$
        \EndIf
        \If{$!cond.values().contains(iter)$} \Comment{Unsupported condition}
            \State Return $false$
        \EndIf
        \State $compOp \leftarrow cond.function()$ \Comment{Compare operator}
        \If{$!compOp.oneOf(\{<, ~\le, ~>, ~\ge \})$}
            \State Return $false$ \Comment{Unsupported condition}
        \EndIf
        \State $ub \leftarrow cond.use.extractUpperBound(iter)$
        \If{$upd.rhs.type() \neq AddExpr$} \Comment{Linear, positive $inc$}
            \State Return $false$
        \EndIf
        \State $inc \leftarrow upd.rhs.extractIncrement(iter)$
        \If{$!IntConst.All(iter,~lb,~inc)$} \Comment{Required constants}
            \State Return $false$
        \EndIf
        \If{$isAssigned(iter, l.statements() \setminus upd)$} \Comment{Iteration variable modified}
            \State Return $false$
        \EndIf
        \If{$l.hasBreak()$} \Comment{$break$ statement in loop}
            \State Return $false$
        \EndIf
        \State Return $iter, ~lb, ~ub, ~inc, ~init, ~upd$
    \EndProcedure
    \end{algorithmic}
\end{algorithm}

Soot provides a toolkit that returns a list of the loops inside a given function. We pass each such loop $l$ to the procedure $isCanonical$. If $isCanonical(l)$ returns $false$, the loop~$l$ is ignored and not parallelized. If the loop is indeed canonical, we extract the iteration variable (iter), lower bound (lb), upper bound (ub), increment (inc), the init statement (init) and the update statement (upd), and store the same for later use.

\subsection{Variable scoping}
\label{ss:scope}
Local variables and non-local variables need to be handled separately for dependence analysis. In a parallel runtime, variables in the loop that are locally scoped are private and allocated on the stack of the running thread. On the other hand, non-local variables are shared across threads. Therefore, we must be able to differentiate local variables from their non-local counterparts. However, unfortunately, Java bytecode does not contain information about variable scopes.

\begin{algorithm}
    \footnotesize
    \caption{\label{algo:scope}{Algorithm to generate the set of local variables.}}
    \begin{algorithmic}[1]
    \Procedure{isLocal}{$var, ~init, ~upd$}
        \If{$var.name.startsWith("\$")$} \Comment{Jimple temporary}
            \State Return $true$
        \EndIf
        \State $initBCI \leftarrow init.getBCI()$ \Comment{BCI: Bytecode index}
        \State $endBCI \leftarrow upd.getBCI() + 3$ \Comment{$upd$ is 3 bytes}
        \State $varStartBCI \leftarrow var.getIndexBCI()$ \Comment {LocalVariableTable}
        \State $varEndBCI \leftarrow varStartBCI+var.getLength()$
        \If{$varBCI \ge initBCI ~AND ~varBCI < endBCI$}
            \State Return $true$
        \EndIf
        \State Return $false$
    \EndProcedure

    \Procedure{getLocalVars}{$body, ~init, ~upd$} \Comment {$body$: soot.Body}
        \State $vars \leftarrow body.getLocals()$
        \State $localVars \leftarrow \emptyset$
        \While{$!vars.empty()$}
            \State $var \leftarrow vars.get(0)$
            \If{$\textsc{IsLocal}(var, ~init, ~upd)$}
                \State $localVars.insert(var)$
            \EndIf
            \State $vars.remove(0)$
        \EndWhile
        \State Return $localVars$
    \EndProcedure
    \end{algorithmic}
\end{algorithm}

Scoping information can be derived indirectly from the local variable table in the bytecode, if the source code is compiled with the {\tt -g} flag. For each variable in the function, the local-variable-table entry in the class file contains the bytecode index of its initialization, its size, and the length in bytes for which it stays live. We directly use ASM (the front-end used by Soot) to read the class file and return the local-variable-table entry corresponding to each variable. Procedure $isLocal$ in Algorithm \ref{algo:scope} checks if the given $var$ is local. A variable is local to the loop if and only if its bytecode index and length are within the bounds of the bytecode indices of $init$ and $upd$. Jimple's temporary variables prefixed with a {\tt \$} are always considered local to the loop. For each canonical loop, we generate the set of local variables using the procedure $getLocalVars$ in Algorithm \ref{algo:scope}.

\subsection{Dependence Analysis on Scalars and Fields}
\label{ss:scalardep}

Being able to discriminate between local and non-local variables in the loop, along with the fact that local variables can be accommodated on the thread's local stack on parallelization, helps parallelize more loops when compared to the conservative case. Figure \ref{f:scopeCode} shows a loop writing to a local as well as a non-local variable. As the variable {\tt c} at line $4$ can be allocated on the local stack of the thread on parallelization, the iterations of the loop running in parallel will be writing to different memory locations. Therefore, this would not constitute a dependence. On the other hand, as the variable {\tt a} is declared outside the loop, each iteration of the loop running in parallel would write to the same location. Therefore, this would indeed constitute a write-after-write dependence. Assuming that we omit line $5$ from the code, we can parallelize the loop because it would not contain any dependence. But even in this case, if we were not able to discriminate between the scopes of variables, we would not parallelize the loop. Also note that object fields can be considered special non-local variables.

\begin{figure}
    {
\begin{lstlisting}
public void f(int n) {
    int a = 0;
    for (int i=0; i<n; i++) {
        int c = i;
        a = i * 2;
    } }
\end{lstlisting}}
    \caption{A loop containing a write to a local variable {\tt c} and to a non-local variable {\tt a}.
    }
    \label{f:scopeCode}
\end{figure}

To identify dependence for scalar variables or object fields, we only need to check for writes to variables inside the loop. There would be a write-after-write dependence every time a non-local variable or an object field is written to inside the loop. Specifically, we iterate over each statement of the given loop. For each definition statement, we check if a field reference is being written to. If yes, the loop is rejected for parallelization. Otherwise, the variable being written to is looked up in $localVars$. If it is not found, \tool{} has identified a dependence and the loop is not parallelized. If no such dependence is detected after processing all the statements, the loop is passed on to the next module.

\subsection{Dependence Solver for Array References}
\label{ss:constraints}

In order to safely mark a loop as parallelizable, for each write to an element of an array, it is crucial to make sure that the same element is not accessed (either read or written) in any other iteration of the loop. Analyzing array references for dependence is difficult because the runtime values of the array indices are not available during static analysis. We propose to formulate this dependence analysis as a {\em satisfiability problem}. The Z3 Theorem Prover can solve the satisfiability problem if given a set of constraints, and returns a satisfying assignment, if it exists. Hence we generate constraints based on program logic, loop iteration variable and array indices, and feed them to Z3. We pose the satisfiability problem as follows: "Is there a satisfying assignment for the program variables under the given constraints, such that the indices of the array references being observed can be the same for two different iterations of the loop?"

To make the array-references easier to work with, we separate them into three sets, namely {\tt arReads} (set of array-read references), {\tt arWrites} (set of array-write references) and {\tt arRefs} ($arWrites \cup arReads$). Dependence analysis for arrays is then done in the following three phases: (i) alias analysis, (ii) constraint generation, and (iii) running Z3.

Only those references that point to the same array object can have a mutual dependence. Therefore, \textit{alias analysis} is crucial for improving the precision of the dependence analysis. We generate the constraints for pairs of elements from $arWrites$ and $arRefs$ if and only if their array objects alias. Here, we identify the aliases using the \textit{GeomPTA} and \textit{Spark} pointer analyses provided by the Soot framework.

To identify if there is a dependence between the given pair of aliasing array references, the minimum constraint is that the array reference indices should not be equal over two different iterations. However, this would be meaningless without capturing the program states in logic; each variable may be assigned results from complex computations. Considering all these factors, we now describe how do we generate the set of constraints; see Algorithm~3. Note that we limit the types of values on the right-hand side of the assignment statements that occur in the {\it def-use chain} of the indices to \texttt{IntConstant}, \texttt{Local}, \texttt{JAddExpr}, \texttt{JMulExpr}, and \texttt{JSubExpr}, denoting common integer-arithmetic expressions for forming indices in parallel programs. From now on, we use $op \in \{+, ~-, ~*\}$ to represent a supported operator.

\begin{algorithm}
    \footnotesize
    \caption{\label{algo:constraints}{Algorithm to generate constraints.}}
    \begin{algorithmic}[1]
    \Procedure{N}{$y, ~i$}
        \If{$y \in localVars$}
            \State Return $y^i$ \Comment{$y$ in iteration $i$}
        \EndIf
        \State Return $y$
    \EndProcedure

    \Procedure{stmtC}{$s, ~i, ~iterVs, ~lbs, ~ubs, ~l$}
        \If{$s$ is $IdentityStmt$} \Comment{The values come from Parameters}
            \State Return $true$
        \ElsIf{$s: ~y \leftarrow (...) ~AND ~y \in iterVs$} \Comment{Other loop's $iter$}
            \State $lbCur \leftarrow N(y, ~i) \geq ~lbs[y]$
            \State $ubCur \leftarrow N(y, ~i) < N(ubs[y], ~i)$
            \State $cU \leftarrow \bigvee_{d \in Def(ubs[y], ~l.head)} \textsc{stmtC}(d, ~i, ~iterVs, ~lbs, ~ubs, ~l)$ 
            \State Return  $lbCur ~AND ~ubCur ~AND ~cU$
        \ElsIf{$s: ~y \leftarrow k ~AND ~k$ is $IntConst$}
            \State Return $N(y, ~i) == k$
        \ElsIf{$s: ~y \leftarrow x_1 ~op ~x_2 ~AND ~x_1, ~x_2$ are scalars}
            \State $cx_1 \leftarrow \bigvee_{d \in Def(x_1,s)} \textsc{stmtC}(d, ~i, ~lbs, ~ubs, ~l)$
            \State $cx_2 \leftarrow \bigvee_{d \in Def(x_2,s)} \textsc{stmtC}(d, ~i, ~lbs, ~ubs, ~l)$
            \State Return $N(y, ~i) == (N(x_1, ~i) ~op ~N(x_2, ~i)) \wedge cx_1 \wedge cx_2$
        \EndIf
    \EndProcedure

    \Procedure{depC}{$w, ~r, ~l, ~iterVs, ~lbs, ~ubs$}
        \State $c_1 \leftarrow \bigvee_{d \in Def(w.index, ~w.stmt)} \textsc{stmtC}(d, ~0, \emptyset, ~iterVs, ~lbs, ~ubs, ~l)$ 
        \State $c_2 \leftarrow \bigvee_{d \in Def(r.index, ~r.stmt)} \textsc{stmtC}(d, ~1, \emptyset, ~iterVs, ~lbs, ~ubs, ~l)$ 
        \State $lc \leftarrow ~l.iter^0 ~!= ~l.iter^1$  \Comment{Different iterations of loop}
        \State $c_{dep} \leftarrow \textsc{N}(w.index) ~== \textsc{N}(r.index)$ \Comment{The dependence}
        \State Return $c_1 \wedge c_2 \wedge lc \wedge c_{dep}$
    \EndProcedure

    \Procedure{loopC}{$arWrites, ~arRefs, ~l, ~iterVs, ~lbs, ~ubs$}
        \State  Return $\bigvee_{w \in arWrites} \bigvee_{r \in arRefs} \textsc{depC}(w, ~r, ~l, ~iterVs, ~lbs, ~ubs)$
    \EndProcedure
    \end{algorithmic}
\end{algorithm}

The procedure \textsc{LoopC} in Algorithm \ref{algo:constraints} generates the set of array dependence constraints for each pair of elements from $arWrites$ and $arRefs$ in a given loop $l$, the set of iteration variables, lower bounds and upper bounds of all the loops of the function $iterVars$, $lbs$ and $ubs$, respectively. The constraints are encoded such that if the set of constraints is \textit{satisfiable}, there is a dependence.
The procedure \textsc{N} returns a mapping from the set of program variables to a set of logical variables.
If a variable $y$ is non-local to the loop, there is an identity mapping. If the variable is local to the loop, $y^i$ denotes the logical variable for the program variable $y$ in the $i^{th}$ iteration.
The procedure \textsc{StmtC} recursively generates the constraints for the program logic, starting at a given statement.
If the value of the variable comes from a parameter, we stop and return identity constraint $true$.
If the variable is the iteration variable of another (nested) loop, we constrain it to its lower-bound and upper bound.
Lines 13-14 suggest that if the value assigned is an integer constant, the variable should be constrained to that integer value.
Finally, lines 15-18 recursively generate the constraints for the operands on the right hand side.
Procedure \textsc{DepC} generates the constraints for a given pair of elements from the sets $arWrites$ and $arRefs$.
Lines 22-23 enforce separate iterations and that the indices of the references be equal for a dependence.
This procedure is called from \textsc{LoopC}, which generates the entire constraint set by taking the disjunction of the dependence constraints for each pair of $w \in arWrites$ and $r \in arRefs$.

In the absence of the value of a variable during static analysis, the constraints would be weaker, and hence, easier to satisfy. Therefore, \tool{} takes a conservative approach in the absence of surety, maintaining soundness.

\begin{figure}
    \begin{center}
    \begin{tabular}{c}
\begin{lstlisting}
public void foo(int ar[]) {
    for(int i=0; i<10000; i++) {
        int k1 = f1(i);
        int k2 = f2(i, k1);
        int k3 = f3(i, k2);
        ar[k3] = k2; } }
\end{lstlisting}\hspace*{\fill}
    \end{tabular}
    \end{center}
    \caption{A loop containing an array write.
    }
    \label{fig:logicCode}
    \end{figure}

Figure \ref{fig:logicCode} shows a simple loop given to the program for dependence analysis. $k1$, $k2$ and $k3$ are locally-scoped, and $f1$, $f2$ and $f3$ can be one of the supported operators as mentioned above. Equation~\ref{eq:example} shows the constraints generated for the array reference at line $6$ with itself, for two separate iterations represented by the superscript:
\begin{multline} 
    \label{eq:example}
    (k3^u == f3(i^u, k2^u)) \wedge (k2^u == f2(i^u, k1^u)) \wedge \\ 
    (k1^u == f1(i^u)) \wedge (i^u \geq 0) \wedge (i^u < 10000) \wedge \\
    (k3^v == f3(i^v, k2^v)) \wedge (k2^v == f2(i^v, k1^v)) \wedge \\
    (k1^v == f1(i^v)) \wedge (i^v \geq 0) \wedge (i^v < 10000) \wedge \\
    (i^u \neq i^v) \wedge (k3^u == k3^v)
\end{multline}

The generated constraints are passed to the Z3 $Solver$. If the $Solver$ returns $Status.UNSATISFIABLE$, the indices in the array references cannot be equal in different iterations, thus deeming that the loop is free of dependencies and can be parallelized. Otherwise, the $Solver$ was able to satisfy the constraints and the loop contains some dependence; the loop is not parallelizable in such a case.

The amount of computation and memory required for the points-to-analysis and solving the constraints using Z3 enforces that the analyses in \tool{} must not be done at runtime. Even though we would have much more information about the loops during JIT compilation, we would incur a high overhead on each run. This is the main reason we designed \tool{} as a static analysis.

\subsection{Supporting Calls to Pure Functions}
\label{ss:purity}

While dependences caused by scalars, fields and array references inside the loop body are taken care of by the analyses presented in preceding sections, analyzing function invocations inside loops is not straightforward. Such an analysis requires checking the variables and references in the invoked functions for dependences in context of the loop, in some sense, extending the above-mentioned analyses to external functions. This problem can be solved by checking if the function has references to non-local objects, i.e. if the function is {\em pure}, since purity will ensure that there are no dependences among different iterations of the loop, thus enabling the parallelization of loops containing function calls.

Soot already provides an in-built purity analysis module, but in our preliminary testing we found it to be generating incorrect results for recent Java versions. Hence we have written a new interprocedural purity analysis that uses Soot's points-to analysis~\cite{spark} and call graph construction modules. Our purity analysis module analyzes the functions called within canonical {\tt for} loops for purity, and provides the results in terms of two parameters: {\em read impurity} and {\em write impurity}. Read impurity implies that the function only accesses or reads non-local entities but does not modify the same, whereas write impurity implies that the called function mutates or writes to a non-local entity.

A pure function should not modify the state that existed before its invocation~\cite{purity}. A Java method can do so through static fields and variables, references passed as parameters, and function calls. 
We first mark functions that access static field references as impure. The remaining function calls are handled using the interprocedural part of the analysis, stating the caller function as also impure if the called function is impure. For the parameters, any object reachable from the parameter should not be accessed inside the function, else the function would be deemed impure.

We use the points-to analysis provided by Soot, which establishes a relationship between the variables and the objects they point to, to get all the objects that are transitively reachable from the objects pointed to by the parameters. We maintain a list that contains all the local variables that can point to an external object. This is done by iteratively updating the list by adding the local variables that can alias with ones already in the list, and then also adding the ones that store the fields of external objects. Whenever any object referenced by a variable from this list is read, it indicates read-impurity, whereas if they are written then it indicates write-impurity. We store these results and use them to determine whether a function called from within a loop is pure or not.
As an example, after extending the dependence analysis with purity analysis, \tool{} can successfully mark the {\tt for} loop in Figure~\ref{fig:intro} as parallelizable.

\section{Runtime Support}
\label{s:runtime}

Performing the analyses shown in Section~\ref{s:analysis} inside {\tt TornadoVM} during runtime would have simplified actually parallelizing the loop. However, performance concerns render such sophisticated analyses in Java virtual machines infeasible, which brings in the problem of communicating the analysis results from static analysis to {\tt TornadoVM}.
In this section, we describe how \tool{} conveys static analysis results to {\tt TornadoVM}, along with the support added in {\tt TornadoVM} to use the conveyed results for loop parallelization.

As {\tt TornadoVM} requires the \parcons{} annotation to be placed above parallelizable {\tt for} loops in the Java source code and Soot works with Java class files, the straightforward solution would have been to insert these annotations in the bytecode itself. We refrained from using this approach because our analysis and parallelization are very sensitive to the bytecode indices and the local variable table generated after static compilation, and there is no one-to-one correspondence among the same between Java source code and bytecode. Even a slight change in the instructions while generating a new class file can be fatal to the program semantics.

The safer solution for communicating the static analysis results of \tool{} to {\tt TornadoVM} runtime is to create a map $AnnotationMap: ~signature \rightarrow List(Annotations)$. Each annotation consists of the $start$, $length$ and $slot$ (in the stack frame) of the iteration variable of the parallelizable loop, looked up from the {\tt LocalVariableTable} in the corresponding Java class file. $signature$ denotes the bytecode signature of a Java method. This $AnnotationMap$ is written to disk after \tool{} returns and supplied to the VM (see Figure~\ref{fig:block-diagram}).
Following is an example $AnnotationMap$ for the code in Figure~\ref{fig:logicCode} with adapted functions {\tt f1}, {\tt f2}, {\tt f3}:\\ $\{<DepTest: foo([I)V> : [{start:2,length:35,slot:1}]\}$.

When TornadoVM encounters a method call, it reads the existing \parcons{} annotations from the classfile and adds all the annotations to a method-local data structure.
We modified TornadoVM to read, in addition to the annotations in the classfile, the $AnnotationMap$ generated by our static analysis, and extend the data structure maintained by TornadoVM by looking up the signature of the called method in the map.
This solution bypasses the need to change the classfile, but is still able to communicate the results to the runtime.

\section{Discussion}
\label{s:discsn}

In this section, we highlight few subtle aspects of the design decisions made while implementing \tool{}, along with a discussion on the reasoning and the alternatives.

{\em 1. Static initializers.}
\tool{} treats calls made to static initializers ({\tt classinit} methods) as impure.
We handle static initializers conservatively because it is difficult to predict when are they called during runtime (first reference to a class); marking them impure should not be an issue because in general they are used to assign values to static fields, which are essentially shared (global) variables anyway, thus leading to impurity for the enclosing loop.


{\em 2. Annotating class files with static-analysis results.}
As mentioned in Section~\ref{s:runtime}, we have chosen to include static-analysis results in separate files (containing the $AnnotationMap$), for simplicity.
In a production scenario, the results can be added as annotations in the class-files themselves, without loss of generality, depending on the correctness of the support to maintain local-variable offsets in Soot and ASM.
We leave this as a future engineering exercise.

{\em 3. Validating static-analysis results.}
For an approach that uses static-analysis results to perform VM-level optimizations without programmer intervention, one of the ways to assess the precision and the usefulness would be to validate the results through a parallel-programming expert.
In this paper, we validate the precision and correctness of \tool{}-results by checking whether the loops identified as parallelizable are a subset of the manually annotated loops by TornadoVM designers, and the usefulness by measuring the speed-ups achieved with the parallelized loops (see Section~\ref{s:eval}).
A future work could be to export \tool{} as an IDE plugin that suggests parallelizable loops to programmers who can either accept or reject the suggestion; we mark this as an interesting software-engineering exercise.

\section{Implementation and Evaluation}
\label{s:eval}

\small
\begin{figure*}
    \begin{tabular}{lllllllll}
        \toprule
        Name               & \# loops        & \# annot        & \# Id         & tAnalysis (s)      & szClassfile (B)       & szAnnotMap (B)      & tParallel (ms)       & tRead (ms)    \\
        \midrule
        Convolution2D      & 2               & 2               & 1             & 1                  & 4527                  & 391                 & 1546.23              & 7.94          \\
        Euler              & 5               & 2               & 1             & 1                  & 4767                  & 497                 & 191.03               & 8.22          \\
        FDTDSolver         & 6               & 6               & 2             & 4                  & 7609                  & 886                 & 429.93               & 7.56          \\
        FlatMapExample     & 2               & 1               & 1             & 1                  & 3815                  & 389                 & 658.05               & 20.72         \\
        GSeidel2D          & 2               & 2               & 0             & 1                  & 4592                  & 0                 & 59.35                & 0             \\
        HilbertMatrix      & 2               & 2               & 1             & 1                  & 3218                  & 390                 & 863.65               & 12.64         \\
        Jacobi1D           & 2               & 2               & 2             & 1                  & 4785                  & 739                 & 418.06               & 7.06          \\
        Jacobi2D           & 4               & 4               & 4             & 1                  & 5222                  & 758                 & 70.66                & 4.45          \\
        Mandelbrot         & 3               & 2               & 0             & 1                  & 9186                  & 0                 & 13740.69             & 0             \\
        MatrixMul2D        & 3               & 2               & 2             & 1                  & 5304                  & 840                 & 50.73                & 25.99         \\
        MatrixTranspose    & 2               & 2               & 2             & 1                  & 4229                  & 391                 & 1031.58              & 25.52         \\
        Montecarlo         & 1               & 1               & 0             & 3                  & 3615                  & 0                 & 418.34               & 0             \\
        Saxpy              & 1               & 1               & 1             & 1                  & 3658                  & 371                 & 842.44               & 12.41         \\
        SGEMMFPGA          & 3               & 2               & 2             & 1                  & 3835                  & 388                 & 1058.39              & 13.16             \\
        \textbf{GeoMean}   & \textbf{2.30}   & \textbf{1.96}   &               & \textbf{1.22}      & \textbf{4647.22}      & \textbf{466.25}     & \textbf{479.48}      &               \\
        \bottomrule
    \end{tabular}
    \caption{Evaluation metrics.
 Out of the total number of loops (loops) and the manually annotated \parcons{} loops (annot), \tool{} identified loops are shown in the Id column.
    tAnalysis denotes the time taken by static analysis in seconds. szClassfile and szAnnotMap respectively denote the size of the benchmark classes and static-analysis results in bytes. tParallel and tRead respectively denote the total execution time and run-time overhead of our approach in milliseconds.
    }
    \label{fig:bm-stats}
\end{figure*}
\normalsize

We have implemented the four static components of \tool{} (highlighted in gray in Figure~\ref{fig:block-diagram}), in the Soot framework~\cite{soot} version 4.1.0, over its Jimple intermediate representation, with different modules implemented independently (thus being candidates for separately useful artefacts as well).
The integration for constraint solving was done with Z3 theorem prover~\cite{z3} version 4.8.11.
We have added the runtime support code to TornadoVM version 0.11, and ran our experiments on an 11th Gen Intel Core i5 machine, bundled with an Intel Iris Xe Graphics chip.

We have evaluated our tool on 14 benchmarks from the PolyBench suite~\cite{polybench}, adapted to Java by the TornadoVM team~\cite{tornadovm} itself.
The parallelizable loops in all these benchmarks are already annotated with {\tt @Parallel} constructs, thus providing us a baseline for evaluating the precision and correctness of the loops identified as parallelizable by \tool{}.
Additionally, we have also evaluated our techniques on a series of synthetic benchmarks, written specifically to test individual cases that \tool{} parallelizes, as well as to illustrate cases which pose challenges for further automatic parallelization.
We plan to release our complete implementation, bundled with all the testcases, to the community as open source.
\tool{} can not only be used as a bundle tool to parallelize loops for TornadoVM, but its individual components (particularly the dependence analysis, the purity analysis, and the Soot-Z3 integration modules) can separately be used to develop various other Soot-based program analyses as well.

We now present an evaluation to study the impact of our tool to support loop parallelization for TornadoVM.
In particular, the next four subsections address the following four research questions, respectively:
\begin{itemize}[left=0.2em,nosep]
    \item {\bf RQ1.} How many of manually parallelized loops are marked as parallelizable by \tool{}?
    \item {\bf RQ2.} Are the overheads with respect to the static-analysis time, the storage for $AnnotationMap$, and the time spent in the VM significant?
    \item {\bf RQ3.} How good are the speedups of \tool{}-marked loops in TornadoVM?
    \item {\bf RQ4.} What are the challenges yet to be handled by future static-analysis guided loop parallelizers?
\end{itemize}

\subsection{\tool{} precision}

Columns 2, 3 and~4 in Figure~\ref{fig:bm-stats} show the number of source-code {\tt for} loops, the number of loops manually annotated with {\tt @Parallel} in the TornadoVM benchmarks, and the number of loops identified as parallelizable by \tool{}, respectively.
Each of the benchmarks contains at least one parallelizable loop, with the maximum number of parallelizable loops being 6 (for {\tt FDTDSolver}).
Out of the 38 loops across the 14 kernels, 31 of them were found to contain the {\tt @Parallel} annotation, whereas \tool{} successfully identified 19 of them to be parallelizable.
Thus, across all the benchmarks, \tool{} is able to successfully identify 61.3\% of the manually parallelized loops.
Recognizing that this identification is done statically without any manual intervention (and as shown later, with negligible runtime overhead), we believe that the precision is reasonable.

In order to validate the correctness of the loops identified as parallelizable by \tool{}, we checked whether the identified loops are a subset of the loops annotated manually with {\tt @Parallel}, and we found that this was indeed the case.
On synthetic testcases designed to test individual features of our analysis, we also found few loops to be marked as parallelizable by \tool{} that were not so straightforward to be parallelized manually.
This supports our hypothesis that a manual identification of parallelizable constructs is error-prone and imprecise, and consequently encourages the use of \tool{} for real-world programs and runtimes.

\begin{figure}
    \includegraphics[width=\linewidth]{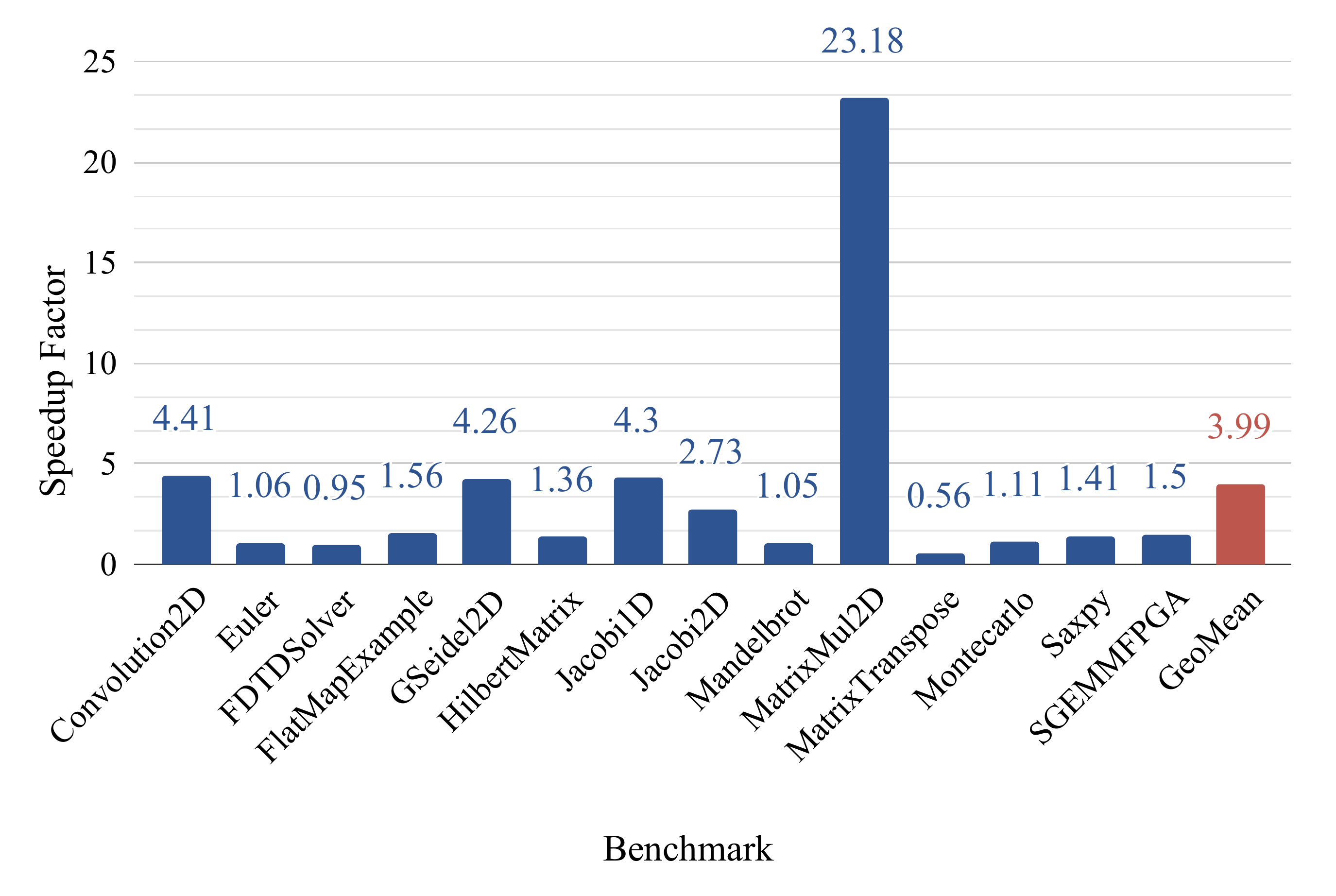}
    \caption{Speed-ups in TornadoVM with \tool{}.
    }
\label{fig:chart}
\end{figure}

\subsection{Time and space overheads}

\tool{} makes use of sophisticated interprocedural points-to and call-graph construction analyses, computed dependence information within loops, purity information about functions, and invokes Z3 from within to check satisfiability of dependence constraints.
Performing all of this during runtime (in a Java VM) would not only be a tremendous engineering effort, but might also be practically infeasible due to the associated analysis cost.
On the other hand, though performing analyses statically takes care of the complexity and practicality to a great extent, our approach incurs additional overhead in terms of conveying the $AnnotationMap$ to, and adding runtime support to read the same in, TornadoVM.
We assess the scale and impact of these overheads next.

Column 5 in Figure~\ref{fig:bm-stats} shows the total analysis time spent by \tool{}, for all the benchmarks.
This includes the time spent by Soot to construct control-flow- and call-graphs, as well as the time spent by Z3 in solving the \tool{}-generated dependence constraints.
We note that the time spent across different benchmarks varies between 1 and 4 seconds; this time is more or less of the same order for different benchmarks due to the similarity in their size, but we expect it to increase proportionally with the size of the benchmark.
Nevertheless, the total analysis time is reasonable to be incurred statically (i.e. offline).

Columns 6 and 7 respectively show the size (in bytes) of the classfiles of various benchmark programs (only the application) and the size of the \tool{}-generated result files.
We observe that the extra space incurred by our approach to convey precise static-analysis results to TornadoVM is very small (on an average 500 bytes), which is just ~10.2\% of the overall class-file size.
This denotes that the results computed by our static analyses are small enough to be conveyed to the VM without much overhead.
As explained in Section~\ref{s:runtime}, we achieve this by storing the results in terms of mostly integer values (storing information in terms of bytecode indices in the class files).

Column 8 shows the total execution time (in milliseconds) of each (parallelized) program under consideration.
Similarly, column 9 shows the time spent by our modified TornadoVM for reading the $AnnotationMap$ and using the results to mark loops as parallelizable.
As can be noted, the runtime overhead of our approach is of the order of a few milliseconds, which is negligible compared to the amount of time that would have been needed for actually performing sophisticated program analyses in the JVM, as well as to the total execution time.
Also note that \tool{} enables program-analysis based parallelization of loops irrespective of the tiered VM component translating the program (interpreter or any of the JIT compiler(s)), whereas even the imprecise version of such analyses could have been performed only if the given method was picked up by a JIT compiler.

\subsection{Achieved speed-ups}

The previous sections assert the precision and the efficiency of the loops parallelized by \tool{}.
We now assess the impact of loop parallelization itself, by comparing the execution times of the sequential and the \tool{}-parallelized versions of various benchmarks, on TornadoVM.

Figure~\ref{fig:chart} shows the speed-ups achieved by the parallel versions of the benchmarks under consideration.
We note that the speed-ups go up to ~23x (for {\tt MatrixMul2D}), and on an average stand at 3.99x.
We also noted slowdowns on few of the benchmarks, specifically {\tt FDTDSolver} and {\tt MatrixTranspose}, and suggest two ways to address the same.
Firstly, the speed-ups may vary if the programs are executed on larger datasets and/or on higher-end systems.
Second and more importantly, the slow-downs indicate that not all of the parallelizable loops are good candidates for parallelization; overheads with respect to communication (particularly TornadoVM's target being heterogeneous systems consisting of GPUs and FPGAs that have high communication overheads, apart from CPUs) is an important factor. We have also observed that the imprecision affects mostly the outer loops (see Section~\ref{ss:further-challenges}), thereby increasing the overall overhead of parallelization when parallelizing inner loops.
We envisage that future studies should adapt the approaches of~\citet{futures} to filter out loops that may not be good candidates for parallelization, for heterogeneous systems.

\subsection{Challenges towards further parallelization}
\label{ss:further-challenges}

The loops that are not parallelized by \tool{} but can be labelled as parallelizable manually may be non-exhaustively categorized into the following two categories.

{\em 1. Library function calls.}
Even though we account for purity of function calls during analysis in \tool{}, library functions are marked (conservatively) as impure.
This means not parallelizing all such loops which even contain functions like {\tt sqrt}; \tool{} failed to parallelize {\tt Montecarlo} because of this reason.
We treated library functions as impure due to two reasons:
First and more important, in a real-world scenario, the JDK installation on the target machine may be different from that available for static analysis, which may lead to invalidity of statically generated results.
Second, including library calls in the analysis blows up the size of Soot's call graph (due to its imprecision), thus making the analysis unsuitable for general-purpose machines with moderate compute capabilities.
Approaches such as the PYE framework~\cite{pye} handle both these challenges by generating results conditional on the library methods, and the same can be incorporated, if deemed suitable for precision and scalability, by integrating \tool{} into the same.
\begin{figure}
    \begin{center}
    \begin{tabular}{c}
    {
\begin{lstlisting}
void hilbert(float[] output, int rows, int cols) {
    for (int i = 0; i < rows; i++) {
        for (int j = 0; j < cols; j++) {
            output[i*rows+j] = 1.0/((i+1)+(j+1)-1); }}}
\end{lstlisting}}
    \end{tabular}
    {\renewcommand{\baselinestretch}{0.77} \caption{\label{f:hilbert}Code for Hilbert Computation.}}
    \end{center}
\end{figure}

{\em 2. Unknown upper bounds during static analysis.}
Multiple loops in our evaluation set are not parallelizable because of the lack of the value of upper bound of the iteration variable during static analysis. As the upper bounds of loops are generally runtime values, the constraints given to the Z3 solver are weaker than they would be at run time.
Figure \ref{f:hilbert} is an example of such a loop; as the values of $rows$ and $cols$ are missing during the analysis, the solver is able to find a dependence when $j >= rows$, for $i_0 = 0$ and $i_1 = 1$.
Hence, the outer loop is not parallelized. In our evaluation, we found that about 83\% of imprecision of \tool{} was due to unknown upper bounds.
In future, we plan to research how can we generate results in terms of conditions on loop bounds that can be efficiently resolved during execution.

\section{Related Work}
\label{s:related}

Automating loop parallelization for achieving performance on multi-core (and recently, heterogeneous) systems has been studied for long~\cite{loopParOptimal1}, and its optimality for even simple loops has been proved to be undecidable~\cite{loopParOptimal2}.
In this section, we primarily focus on related works that propose significant advancements in performing (loop) parallelization using dependence and/or purity analysis.

The precision of finding and solving dependence constraints directly affects the parallelization, and hence several dependence analysis algorithms have been proposed.
Partition-based merging implemented in Parascope~\cite{parascope} works by separating the array into separable minimally-coupled groups.
Merging direction vectors~\cite{automaticProgramParallelization} is a very common dependence analysis algorithm used by various automatic parallelization tools such as Automatic Code Parallelizer~\cite{automaticCodeParallelizer}.
Symbolic test and Banerjee-GCD test~\cite{bannerjee} are used to detect data dependence among array references by assuming the loop in normal form and the loop indices to be affine functions.
TornadoVM in itself does not use dependence analyses, due to the overhead of performing these checks during program execution.
Our dependence analysis is based on Z3, does not require indices to be affine functions (is able to solve non-linear arithmetic), and most interestingly is performed statically (that is, without incurring any analysis or checking overheads during program execution in the JVM).

Constraint solvers such as Z3 have been extensively used for verification of programs.
Bounded model checking~\cite{bmc} is a popular way of finding bugs in programs.
Satisfiability modulo theory has been extended for verification of higher-order programs~\cite{hop}, and multi-threaded program verification~\cite{multithread}.
Constraint satisfiability based techniques have also been used for quantification of information flow in imperative programs using a SAT-based QIF~\cite{qif}.
On the other hand, Pugh and Wonnacott~\cite{constraintArray} were among the first to propose the usage of constraint-solving for dependence analysis.
Inspired by these prior works, in this work, we feed the constraints identified by our dependence analysis written in Soot to the Z3 solver, and marked loops for parallelization where the dependence constraints are satisfied.
To the best of our knowledge, ours is the first approach that integrates Soot with Z3 for this purpose.

Purity analysis~\cite{purity, purepaper} identifies side-effect free functions, and is imperative for parallelization of otherwise dependence-free loops consisting of calls to functions that may be impure.
Süß et al. propose a C extension~\cite{purityLoops} that marks pure function calls to support parallelization of polyhedral loops.
\tool{} implements a stand-alone purity analysis component (which works for recent versions of Java, unlike prior implementations in Soot), and hence naturally supports programs containing function calls inside Java {\tt for} loops.

Few prior works have proposed dependence analysis based loop parallelization with hybrid static+dynamic strategies.
Oancea and Rauchwerger~\cite{hybrid1} use runtime information to improve the performance of static dependence analysis for FORTRAN.
Recently, Jacob et al.~\cite{hybrid2} used staged dependence analysis while parallelizing Python loops on GPUs, to determine loop bounds and variable types that cannot be determined statically (Python being a dynamically typed language).
Thakur and Nandivada~\cite{pye}, though for a different set of analyses, propose a static+JIT approach that statically encodes dependencies between Java application and libraries precisely as conditional values and resolving the same during JIT compilation.
Our approach facilitates loop parallelization on a dynamic Java runtime, by offloading complex program analyses to static time, and can be extended using such hybrid strategies to improve the precision further.

\section{Conclusion}
\label{s:concl}

Loop parallelization, though one of the most promising ways to speed-up programs on multi-core and heterogeneous systems, requires performing several expensive analyses for automation.
For a language like Java, where most of the program analysis happens during runtime in a VM (and thus affects the execution-time of programs directly), performing such analyses for loop parallelization not only presents several challenges in terms of integrating program analyses with constraint solvers, but may often also be prohibitively expensive.
In this paper, we proposed an approach that solves this problem by performing the required analyses statically, and conveying the obtained results to a recent JVM that parallelizes loops for heterogeneous architectures.
Our solution involved generating dependence constraints from Java bytecode, feeding them to a constraint solver, supporting calls to pure functions, generating results in a form that is valid in the Java runtime, and modifying the VM to support static-analysis guided parallelization.
Our exposition describes the design decisions and implementation challenges along with our novel solutions in detail, and our tool \tool{} is composed of several modules that can additionally be used for performing more such analyses in future.


\bibliography{main}

%

\end{document}